\documentclass[prb,twocolumn]{revtex4}
\usepackage{psfig}
\usepackage{graphicx}
\begin{document}
 
\title{
Pseudo-half-metalicity in the double perovskite Sr$_2$CrReO$_6$ 
from density-functional calculations
} 

\author{G. Vaitheeswaran and V. Kanchana}
\affiliation{Max-Planck-Institut f\"ur Festk\"orperforschung,
Heisenbergstrasse 1, 70569 Stuttgart, Germany}
\author{A. Delin}
\affiliation{Materialvetenskap, Brinellv\"agen 23, KTH, SE-10044 Stockholm, Sweden} 

\date{\today}
\begin{abstract}
The electronic structure of the spintronic material Sr$_2$CrReO$_6$
is studied by means of
full-potential linear muffin-tin orbital method.
Scalar relativistic calculations 
predict Sr$_2$CrReO$_6$ to be half-metallic with a magnetic moment 
of 1 $\mu_B$. When spin-orbit coupling is included, the 
half-metallic gap closes into a pseudo-gap, and an unquenched rhenium orbital 
moment appears, resulting in a significant 
increase of the total magnetic moment to 1.28 $\mu_B$.
This moment is significantly larger than the experimental moment 
of 0.9 $\mu_B$. 
A possible explanation of this
discrepancy is that the anti-site disorder in Sr$_2$CrReO$_6$ is
significantly larger than hitherto assumed. 

\end{abstract}

\maketitle

                                 
The family of magnetic oxides with an ordered double perovskite
structure are complex materials with high technological potential
in the area of spin electronics.
Double perovskites have the general formula  A$_{2}$BB'O$_6$, 
where A can be an alkali metal such as strontium, calcium, or barium,
or a lanthanide, and B and B' are transition metals. 
Each transition metal site
is surrounded by an oxygen octahedron (sometimes heavily distorted), 
and the A atoms are situated
in the holes produced by eight adjacent oxygen octahedra.

In 1998, it was discovered that
one such double perovskite, Sr$_{2}$FeMoO$_6$,
possesses intrinsic tunneling-type magnetoresistance
at room temperature --- until then only observed in the 
mixed-valent manganese oxides\cite{manganites} ---
making it a hot candidate material
for spin-electronics applications.\cite{kobayashi1998} 
The physical origin of the magnetoresistance in 
Sr$_{2}$FeMoO$_6$ and in the mixed-valent manganese oxides
is half-metalicity, i.e., the material is an insulator 
in one of the spin channels, but a metal in the other. 
This leads to a complete 
spin polarization at the Fermi level, which in turn results in
strongly spin-dependent scattering of the charge carriers and thus
a possibility to influence the resistance using relatively
weak magnetic fields.
Prerequisites for realizing high-performance devices using these materials
are that the half-metalicity to a high degree is preserved at
ambient temperature, and that high quality thin films
of the material can be grown.

In this letter, we investigate the electronic structure of 
the double perovskite Sr$_2$CrReO$_6$ using density functional theory.
This system is particularly interesting since it exhibits the
hitherto largest Curie temperature $T_c$ of all known double perovskites,
635 K,\cite{kato2002} which is a couple of hundred Kelvin higher or more 
than for Sr$_{2}$FeMoO$_6$\cite{Sarma2000} 
as well as for the mixed-valent manganese oxides. 
Sr$_2$CrReO$_6$ is a metallic ferromagnet with a 
saturation magnetic moment of around 0.9 $\mu_{B}$ per
formula unit. At room temperature, the moment is only slightly reduced to
around 0.8 $\mu_{B}$, 
and high quality thin films of Sr$_2$CrReO$_6$ can be produced
in quite a large temperature window.\cite{kato2002,asano2004}
Thus, this material appears to satisfy important technological criteria.
%
The measured saturation moment is quite well reproduced by 
a simple ionic picture of the Sr$_2$CrReO$_6$ system, although
this model takes neither hybridization nor
orbital moments into account.
In the ionic model, the Cr ions, situated on the B sites, 
have a $3d^3$ configuration, leading to
a moment of $3 \mu_B$ antiferromagnetically coupled to the
neighboring Re ions on the B' sites, with configuration $5d^2$
or $2 \mu_B$ per atom.
In total, this gives a saturation spin moment of $1 \mu_B$ per 
formula unit in the case of perfect ordering.

The present density-functional calculations were performed using 
an all-electron full potential linear muffin-tin orbital
method (FP-LMTO), which has been described in 
detail elsewhere.\cite{wills}
In this method, space is divided into non-overlapping
muffin-tin spheres surrounding the
atoms, and an interstitial region.
Most importantly, this method assumes no shape approximation of the
potential, wave functions, or charge density.
Spin-orbit coupling was included in our calculations.
%
The spherical-harmonic expansion of the 
potential was performed up to $l_{max}=6$, and we used a
double basis so that each orbital is described using two
different kinetic energies in the interstitial region. 
Furthermore, we included several pseudo-core orbitals
in order to further increase accuracy.
Thus, the basis set consisted of the
Sr ($4s$ $5s$ $4p$ $5p$ $4d$),
Cr ($4s$ $3p$ $4p$ $3d$),
Re ($6s$ $5p$ $6p$ $5d$), and
O  ($2s$ $2p$) LMTOs.
%
We performed our calculations using the experimentally
determined structure and atomic positions, 
i.e., the tetragonal structure with space group symmetry $I4/mmm$, 
with cell parameters $a=b=5.52$~\AA, and $c=7.82$\,\AA.\cite{kato2002}.
The radii of the muffin-tin spheres were 2.68a$_0$ for Sr, 2.0 a$_0$ for Cr, 
1.98 a$_0$ for Re and 1.6a$_0$ for O respectively. 
%
The direction of the spin magnetic moment
was chosen to be along the c-axis.
The integration in reciprocal space was
performed using 376 k-points in the irreducible Brillouin zone (BZ),
corresponding to 2744 points in the full BZ
for our self-consistent ground-state calculation.
We tried both the 
local spin density approximation (LSDA)\cite{barth_lda} 
and  the generalized gradient approximation (GGA)\cite{pbe} to
the exchange-correlation functional.
In the following, we will concentrate on our GGA results
with the spin-orbit coupling included.


Our calculations reveal two important features of the
electronic structure of Sr$_2$CrReO$_6$.
First, our calculated total magnetic moment $J_z$,
i.e., the sum of the spin and orbital
moments, at perfect ordering of the Cr and Re atoms, is about 
1.28$\mu_B$ per formula unit when spin-orbit coupling is
included, i.e., 
a 28\% larger  
magnetic moment than the one predicted by the ionic picture.
(see Table\,\ref{table:magnetic_moments}).
Second, we find that Sr$_2$CrReO$_6$ is in fact not a perfect half metal. 
Let us discuss the magnetic moment first.
When spin-orbit coupling is neglected, we find the
total spin moment to be precisely 1 $\mu_B$.
With spin-orbit coupling included, the spin moment increases to
1.1 $\mu_B$.
Our calculated spin and orbital Cr and Re $d$ moments are listed in 
Table\,\ref{table:magnetic_moments}.
Note that a large part of the spin moment is delocalized into the
interstitial region, and therefore the individual 
$d$ spin moments of the individual Cr and Re atoms inside
their respective muffin-tin spheres 
in Table, appear small compared
to the ionic values.
\begin{table}[tb]
\caption{
Calculated spin and orbital $d$ magnetic moments
in $\mu_{B}$/atom for Sr$_2$CrReO$_6$, with and without
spin-orbit coupling (SO) included, and for different 
choices of the exchange-correlation functional.
The total moment includes also the small 
spin and orbital moments present in non-$d$ orbitals, and
the spin moment in the interstitial region. 
\label{table:magnetic_moments}
}
\begin{ruledtabular}
\begin{tabular}{l|ll|ll|l}
              & Spin  &        & Orbital &       & Total   \\
              & Cr    & Re     & Cr      & Re    &         \\ \hline

LSDA          & 2.00  & -0.75  &         &       & 1.00    \\
LSDA+SO       & 2.02  & -0.69  & -0.029  & 0.17  & 1.31  \\
GGA           & 2.24  & -0.90  &         &       & 1.00    \\
GGA+SO        & 2.25  & -0.85  & -0.030  & 0.18  & 1.28  \\
ionic picture & 3     & -2     &         &       & 1       \\

\end{tabular}
\end{ruledtabular}
\end{table}



We find a Re $5d$ orbital moment of around 0.18 $\mu_B$ 
and a total orbital moment of also 0.18 $\mu_B$, calculated by
summing up the orbital moments in all muffin-tin spheres. 
Both Cr and Re have less than half-filled $d$-shells, and therefore
the orbital moment is antiparallel to the spin moment for both species. 
Since the Cr and 
Re spin moments couple in antiparallel, the net results is that
the total orbital moment, dominated by the Re orbital moment, is
parallel to the total spin moment which, in turn, 
is dominated by the Cr spin moment. 
Consequently, the orbital moment has the effect of further increasing the total
magnetic moment in this system.

Thus, when spin-orbit coupling is taken into
account, our total predicted magnetic moment
becomes significantly larger than what experimental studies  
indicate, and also larger than what the ionic 
model of Sr$_2$CrReO$_6$ suggests.\cite{kato2004} 
Evidently, the spin-orbit effect is very important
in this compound, which is hardly surprising since 
even in the case of elemental Re metal, 
the spin-orbit coupling is necessary in order to reproduce
the experimentally observed 
band structure and Fermi surface.\cite{mattheiss}
What is surprising on the contrary is that the experimentally found moment is
so close to the ionic value.
A possible explanation of this paradox is that the anti-site disorder
might be significantly larger in this system than previously assumed.
One way to resolve this issue would be to
determine the individual atomic spin and orbital moments
in this system experimentally
by performing, e.g., X-ray magnetic circular dichroism (XMCD)
experiments. 

We now turn to the question of half-metalicity in
Sr$_2$CrReO$_6$, at perfect ordering. 
In our scalar-relativistic calculations, we
indeed find a gap of 0.7\,eV in the majority 
spin channel 
(see top panel of Figure 1), but when spin-orbit 
coupling is included, the band gap 
disappears and turns into a 
pseudo-gap with a low but 
finite density of states (DOS), see 
bottom panel of Figure 1. 
The carriers at the Fermi level remain highly 
polarized;
$D(\downarrow)/D(\uparrow) \sim 13$, where
$D(\downarrow)$ and $D(\uparrow)$ is the 
density of states at the Fermi level
for minority and majority spin, respectively, 
instead of infinity as would be the case 
for a perfect half metal. 
As a first hint as to why the gap disappears, we note that in Re metal, 
the spin-orbit parameter $\zeta \left( r\right)$ of the $t_{2g}$ states in the
spin-orbit Hamiltonian 
$\hat{H}_{SO}=\zeta \left( r\right) {\bf \hat{l}\cdot \hat{s}}$
is around 0.4 eV\cite{mattheiss}, a number that 
decreases to approximately 0.3 eV in the double perovskite 
due to covalency.\cite{oka}
Thus, the half-metallic gap 
and the spin-orbit parameter are of the same order, 
which makes it plausible that the spin-orbit splitting 
is capable of washing away the gap. 

\begin{figure}
\begin{tabular}{c @{} c}
\includegraphics[bb=100 330 500 740,clip,width=4.3cm]{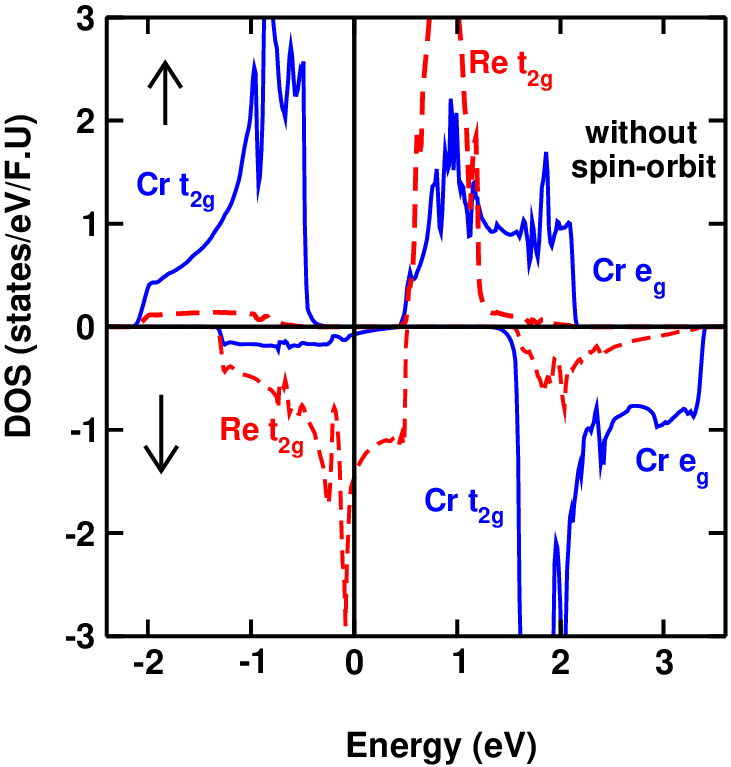} & 
\includegraphics[bb=100 330 500 740,clip,width=4.3cm]{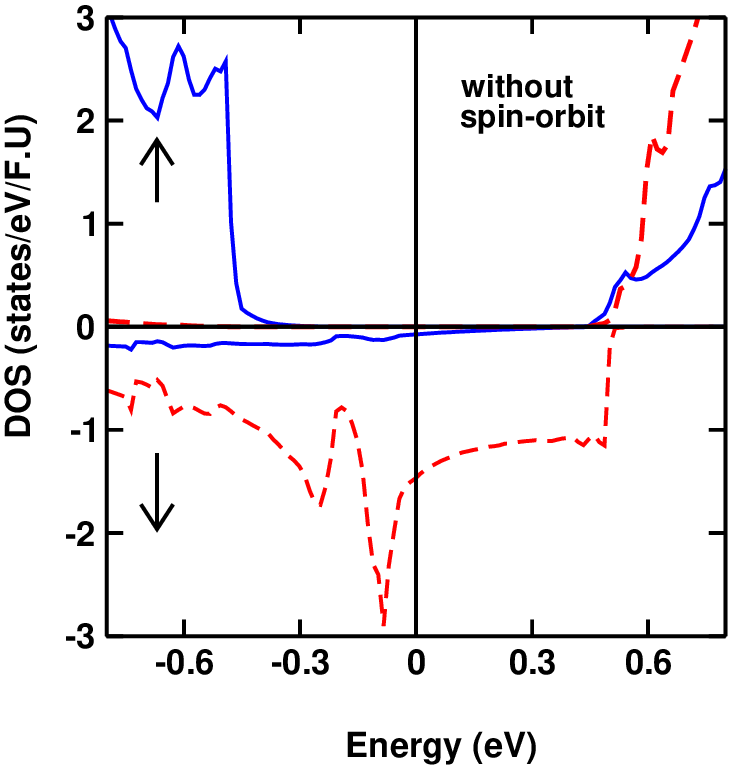} \\
\includegraphics[bb=100 330 500 740,clip,width=4.3cm]{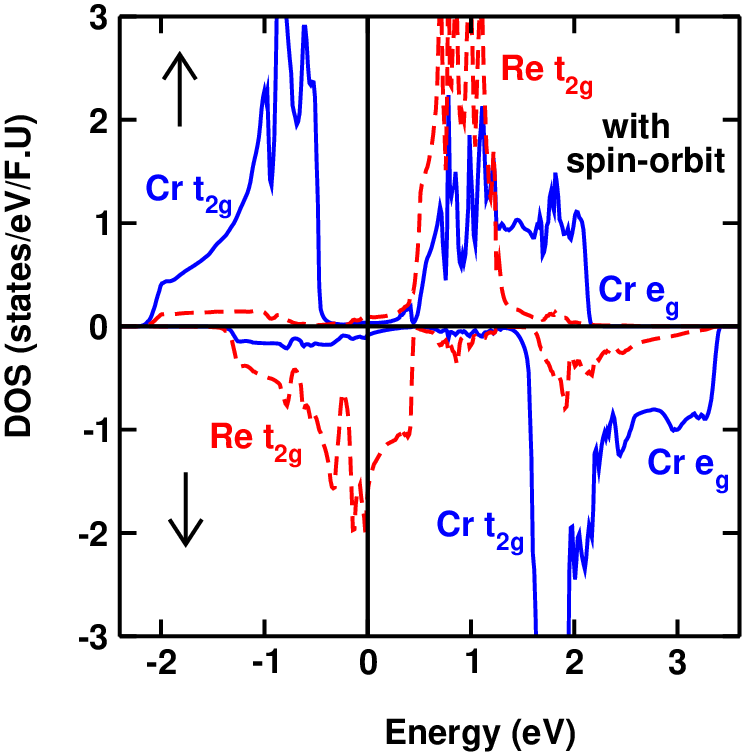} & 
\includegraphics[bb=100 330 500 740,clip,width=4.3cm]{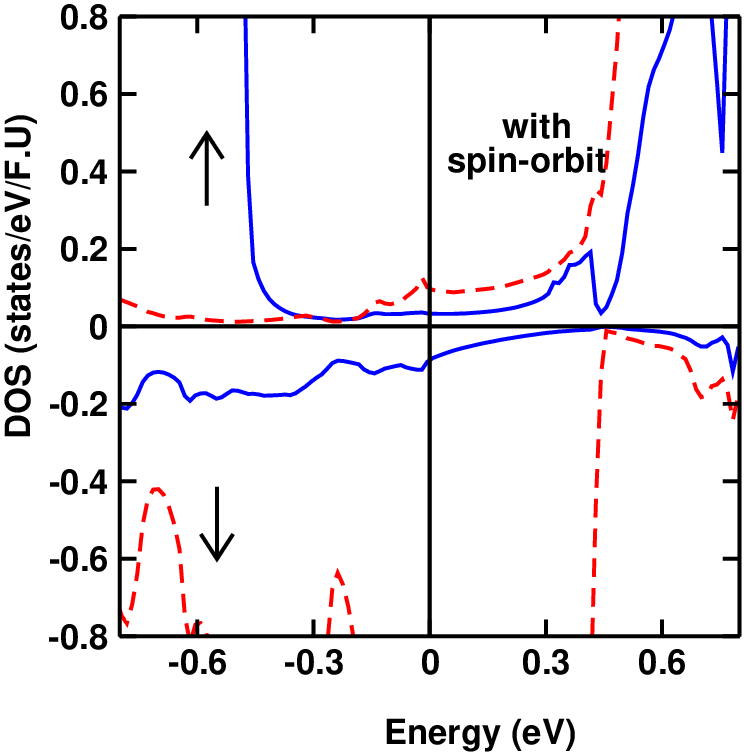}
\end{tabular}
\caption{ 
Orbital-resolved density of states for Sr$_2$CrReO$_6$,
GGA calculation,
without spin-orbit coupling(top panel), and with spin-orbit 
coupling (bottom panel). The blow up of the DOS are given adjacent
The Fermi level is at zero.
}
\label{figure:pdos_fr}
\end{figure}

\clearpage


In order to understand in some more detail why the
half-metalicity is destroyed, we analyze the density of states
of this system, and work out why the spin-orbit coupling
affects the DOS in this particular way.
The basic critical ingredients in the DOS 
are the $d$ states of the Cr and Re atoms, which in turn
are split into $t_{2g}$ and $e_g$ states by the crystal
field produced by the oxygen octahedra, with the $t_{2g}$ states having 
lower energy and place for three electrons per spin channel, whereas the
$e_g$ states are higher in energy and have 
place for two electrons per spin channel. 
In the absence of spin-orbit coupling, 
the $t_{2g}$ and $e_g$ states are eigenstates to the Hamiltonian
and do not hybridize with each other. Similarly, the spin channels do
not hybridize with each other.
(To be exact, the slight distortion of the oxygen octahedra
introduces some very small extra splitting. We will however neglect this in the
following.)
%
In the scalar-relativistic DOS, 
the three-fold degenerate Cr $t_{2g}$ states of the majority spin channel 
are filled. Therefore, the Fermi level ends up in the crystal-field
gap between the Cr $t_{2g}$ and $e_g$ states. A similar situation is 
seen in Sr$_2$CrWO$_6$.\cite{jeng,philipp}
%
Due to the antiferromagnetic coupling of Cr and Re,
it is the minority spin channel in Re which is the occupied one, 
and it contains two electrons.  This means that the Re minority spin
$t_{2g}$ states are only filled to about two-thirds, 
resulting in a high DOS at the Fermi level
in the minority spin channel. 
Due to hybridization, also
the minority spin Cr $t_{2g}$ states obtain a small occupation.
%
In the majority spin channel, the Re $d$ states are essentially
empty; hybridization with the majority Cr $t_{2g}$ states results 
nevertheless in a finite, small occupation.
When spin-orbit coupling is included, the $e_g$ and $t_{2g}$
states are no longer eigenstates to the Hamiltonian, and they will
therefore mix, as will the spin states.
As a result, the high Re $t_{2g}$ DOS at the 
Fermi level in the minority spin channel induces states in
the majority spin channel. 
Since the spin-orbit parameter and the gap are both of the order of
a few tenths of an eV, the result is that
the half-metallic gap closes and the
Cr $t_{2g}$ and $e_g$ peaks become connected. 
The induced pseudo-gap states have Re $t_{2g}$ Cr $e_g$, as well 
as Cr $t_{2g}$ character. 
%
%
In summary, we have analyzed the electronic structure 
of the double perovskite Sr$_2$CrReO$_6$. 
The effect of spin-orbit coupling results in a rather large 
Re orbital moment and as a result, a total magnetic moment
of 1.28 $\mu_B$, whereas our predicted 
scalar-relativistic spin-only moment is precisely 1.0 $\mu_B$. 
Furthermore, the large spin-orbit coupling in Re 
produces a non-vanishing DOS
at the Fermi level in the majority spin channel, destroying the
half-metalicity even at perfect ordering of the Cr and Re sites.
\acknowledgments
The authors acknowledge Prof. O. K. Andersen for his valuable 
suggestions and critical reading of the manuscript.
G.V and V.K acknowledges Max-Planck Society for the 
financial support.
A.D. acknowledges financial support from 
Vetenskapsr{\aa}det (the Swedish Science Foundation).
Dr. H. Kato is acknowledged for sharing
experimental details with us. J.M. Wills is acknowledged for letting us to use his
FP-LMTO code.


%

\begin{thebibliography}{99}


\bibitem{manganites}
J. Z. Sun, W. J. Gallagher, P. R. Duncombe, L. Krusin-Elbaum, R. A. Altman, 
A. Gupta, YuLu, G. O. Gong, and Gand Xiao, 
Appl. Phys. Lett {\bf 69}, 3266 (1996).

\bibitem{kobayashi1998}
K. I. Kobayashi, T. Kimura, H. Sawada, K. Terakura and Y. Tokura,
Nature (London) {\bf 395}, 677 (1998).

\bibitem{kato2002}
H. Kato, T. Okuda, Y. Okimoto, Y. Tomioka, Y. Takenoya, A. Ohkubo, M. Kawasaki
and Y. Tokura, 
Appl. Phys. Lett. {\bf 81}, 328 (2002).

\bibitem{Sarma2000}
D. D. Sarma, P. Mahadevan, T. Saha-Dasgupta, S. Ray, and A. Kumar
Phys. Rev. Lett. {\bf 85}, 2549 (2000).

\bibitem{asano2004}
H. Asano, N. Kozuka, A. Tsuzuki, and M. Matsui,
Appl. Phys. Lett. {\bf 85}, 263 (2004).


\bibitem{wills}
J. M. Wills, O. Eriksson, M. Alouani and O. L. Price in
{\it Electronic Structure and Physical Properties of Solids},
edited by H. Dreyss\'e (Springer, Berlin 2000).


\bibitem{barth_lda}
U. von Barth, L. Hedin, 
J. Phys. C {\bf 5}, 1629 (1972).

\bibitem{pbe}
J. P. Perdew, K. Burke, and M. Ernzerhof,
Phys. Rev. Lett. {\bf 77}, 3865 (1996).


\bibitem{kato2004}
H. Kato, T. Okuda, Y. Okimoto, Y. Tomioka, K. Oikawa,
T. Kamiyama, and Y. Tokura, 
Phys. Rev. B {\bf 69}, 184412 (2004).

\bibitem{mattheiss}
L. F. Mattheiss, Phys. Rev. {\bf 151}, 450 (1966).

\bibitem{oka}
O. K. Andersen (private communication).

\bibitem{jeng}
H. T. Jeng, G. Y. Guo, 
Phys. Rev. B {\bf 67}, 094438 (2003).

\bibitem{philipp}
J. B. Philipp, P. Majewski, L. Alff, A. Erb, 
R. Gross, T. Graf, M. S. Brandt, J. Simon, T. Walther, W. Mader
D. Topwal, and D. D. Sarma, 
Phys. Rev. B {\bf 68}, 144431 (2003).




\end{thebibliography}
 \end{document}